\documentclass[final,1p,times]{elsarticle}
\usepackage{txfonts}
\usepackage{graphicx}
\usepackage{epsfig}
\usepackage{graphics}
\usepackage{amssymb}
\usepackage{subfig}
 \usepackage{amsthm}
\usepackage{amscd}
\usepackage{amsmath}
\usepackage{amsfonts}
\usepackage{amssymb}
\usepackage{graphicx}

\numberwithin{equation}{section}

\newcommand{\be}{\begin{equation}}

\newcommand{\ee}{\end{equation}}
\newcommand{\bea}{\begin{eqnarray}}
\newcommand{\eea}{\end{eqnarray}}
\newcommand{\bna}{\begin{eqnarray*}}
\newcommand{\ena}{\end{eqnarray*}}

\journal{$\ast\ast\ast$}

\begin{document}

\begin{frontmatter}

\title{Mortality cohort effect detection and measurement based on differential geometry}
\author[label1]{Zhang Ning}
 \ead[label1]{nzhang@amss.ac.cn}
\address[label1]{China Institute for Actuarial Science,
 Central University of Finance and Economics, Beijing 100081, P. R. China}
\author[label2,label3]{Zhao Liang }
 \ead[label2]{ liangzhao@bnu.edu.cn}
\address[label2]{Corresponding author, School of Mathematical Sciences,
 Beijing Normal
University, Beijing 100875, P. R. China}
\address[label3]{Laboratory of Mathematics and
Complex Systems, Ministry of Education,
Beijing 100875, P. R. China}

\begin{abstract}
This paper analyzes mortality cohort effect of birth year and develops an approach to identify and measure cohort effects in mortality data set.
The approach is based on differential geometry and leads to an explicit result which can describe how strong the cohort effect is in any period.
This quantitative measurement provides a possibility to compare cohort effects among different countries or groups.  The paper also suggests to use coefficient of variation as a measurement of the whole cohort effect for one country. Data of several countries are taken as examples to explain our approach and results.

\end{abstract}


\end{frontmatter}

\section{Introduction}

      Cohort effect is used in social science to describe variations among individuals who are defined by some shared temporal experience or common life experience, such as year of birth, or year of exposure to radiation (WikiPedia). Many papers and reports have already noted cohort effects  in the mortality data  and also highlighted the existence of cohort effects in different countries \cite{1,2,3}. Among them, it is well known that people born in the U.K. between 1925 and 1945 have experienced more rapid improvement in mortality than generations born in other periods (GAD 1995,2001). In other words, this generation has experienced stronger cohort effect than others. There is also some further research on empirical analysis about the influence of cohort effect on mortality \cite{2,3}, but how much is the cohort effect for a specific group? Weak or Strong? Can we compare cohort effects between two groups? Whether or not should we consider this effect in the process to measure longevity? It is considerable to measure cohort effect for a mortality data set before use. In this paper, we deal with these problems and develop a method to detect and measure cohort effect.

      Since cohort effect has become an important factor in modeling or analysing longevity risk, the method improves the related works to be simpler or more effectively. This conclusion can be drawn from the history of longevity risk's research. Meanwhile, one of our motivations is also inspired by this developing process in the following.

      The enormous improvement in life expectancy is certainly one of the greatest achievements of modern civilization, but unanticipated mortality improvements can pose huge problems to individuals, corporations and governments. In fact, longevity risk, that is, the uncertainty associated with future mortality improvements, has become a prominent risk in many countries and the financial burden to individuals, corporations and governments is also triggered by mortality improvement. Many models have been promoted to acquire information about longevity risk from mortality data set. Among these models, Lee-Carter model is well-known and its derivative models have showed their success in forecasting future mortality.

In Lee-Carter model, mortality can vary across individuals as they age (aging or life cycle effects) and time goes (period effects). Let $\mu_x(t)$ denotes the central death rate for age x at time t, the model can be expressed as the following log-bilinear form \cite{4}:
$$ \ln \mu_x(t)=\alpha_x +\beta_x k_t +\epsilon_{x,t}.$$

Here the change in the level of mortality over time is described by the mortality index $k_t$, $\alpha_x$ describes the age-pattern of mortality averaged over time and $\beta_x$ describes the deviation from the averaged pattern when $k_t$ varies. Finally, the quantity $\epsilon_{x,t}$ denotes the error term, with mean $0$ and variance $\sigma^2$.

The Lee-Carter model can capture aging effects and period effects. To make it perfect, possible improvements of Lee-Carter model are promoted, which use a random variable $D_x(t)$ for the number of deaths. Here $D_x(t)\sim {\text Poisson}(ETR_x(t),\mu_x(t))$ \cite{7}, $ETR_x(t)$ is the central number of exposed to risk which refers to the total number of person-years in a population over a calendar year in the actuarial literature and $\mu_x(t)$ is given by:
      $$\ln\mu_x(t)=\alpha _x +\beta_x k_t.$$

   The Lee-Carter model generally considers both aging effects and period effects. But the group born in the same year should share some characters which would impact its mortality. To reflect this impact, we had better consider cohort effect in the model. In fact, there are several descriptive works about the existence of cohort effect in some countries. A simple model considering cohort effect is just like this form \cite{3}:
    $$\ln\mu_x (t)=m+\alpha_x+\beta_t+\gamma_{t-x}.$$
    Here $\alpha_x$ denotes the age effect, $\beta_t$ denotes the period effect, and $\gamma_{t-x}$ denotes the cohort effect.

   Renshaw and Haberman also introduced cohort effect into Lee-Carter model in a quantitative way. Their model is the following age-period-cohort (APC) version of Lee-Carter model\cite {8,9}:
     $$ \ln \mu_x (t)=\alpha _x +\beta_x^{(0)}i_{t-x}+\beta_x^{(1)}k_t.$$
     Here an extra item $\beta_x^{(0)}i_{t-x}$ is introduced in order to represent cohort effect.  The model can give rise to two sub-structure models: setting $\beta_x^{(0)}=0$ to get the classic Lee-Carter model and setting $\beta_x^{(1)}=0$ to get age-cohort model. There is still a slightly modified version of the above form\cite{3}:

     But the well-documented problem with these models is that only two of these effects can be identified. Age (years since birth), period (year), and cohort (year of birth) are exact linear functions of each other: Age$=$Period$-$Cohort (introduction of this issue, see \cite{10}, Winship and Harding, 2008). In other words, we can find no unique set of parameters resulting in an optimal fit because of this trivial relation.

     Now, we can make a summary which inspires us to find a way to detect and measure cohort effect:

     1) The cohort effect of birth year exists in some mortality data sets and even we can observe a data set directly to find some evidences. But we do not know the exact information about cohort effect when given any mortality data set.

     2) The cohort effect is added into the Lee-Carter model but this brings some troubles in the calibration of the model. We do not know whether or not this modification can bring any modeling or forecasting improvement for a mortality data set.

     3) We can not precisely understand the reason for these so called cohort effect. But if we know more information about the existence of cohort effect, it will be easier to analyze the causes, such as war, smoking, cultural background or others.

     The facts above inspire us to detect and measure cohort effect . The consequences of the work include the followings at least:

     1) Get exact information about cohort effect for any mortality data sets, which is an important population character for countries.

     2) Decide whether or not to use the APC model, or in other word, whether or not to take cohort effect into account when modeling the mortality data set. Moreover, we can apply our results in analyzing longevity risk to set up a more effective parameter to express the influence of cohort effect.

     3) Analyze underlying causes via the distribution of cohort effect in different periods and countries.

     Here is an example of Chinese mortality for explanation. As we know, when modeling and forecasting mortality, different models will lead to different results which may impact supervision and decision-making. We use Lee-Carter model, Poisson bi-linear model and Mortality decomposition model to analyze Chinese longevity risk respectively. For female age 60, the life remaining expectation in 2020 has different forecasting results: $23.6$ by Lee-Carter model, $27.18$ by Poisson bi-linear model and $27.16$ by mortality decomposition model \cite{17,18}. We also find cohort effect will impact the age-specific and period-specific characters. It leads to different results when estimating error for longevity risk because period effect is impacted by cohort effect.

     If cohort effect can be recognized before apply mortality models to mortality matrix, we can decide whether or not cohort effect should be considered. Furthermore, if cohort effect can be measured by some index, maybe called cohort effect index, we can apply it into mortality models based on statistical methods. In particular, cohort effect index can provide more information about longevity than $\gamma_{t-x}$ dose in APC model. Moreover, APC model uses the linear model to capture cohort effect ( Log function can provide some non-linear information) . But obviously, mortality data is non-linear and looks like a uneven surface. Figure 2 shows the surface of US mortality. It is natural to capture the characters of the data set by analyzing the geometric properties of the corresponding surface. This is the motivation of our theoretical model and we use some techniques in differential geometry to detect and measure cohort effect.

     We organize this paper as follows: In Section 2, we propose the model based on the theory of differential geometry, which attempts to identify cohort effect of any mortality data set. We also give a definition of cohort effect index(CEI) which can illustrate the strength of cohort effect in any period. In Section 3 we show some applications of the model to the mortality data sets of several countries. In Section 4, we define the aggregating index of cohort effect which can represent the strength of the cohort effect for the whole population. Section 5 outlines some potential applications and further developments of this paper.

\section{Theoretical model of detecting mortality cohort effect}

To comprehend the cohort effects of mortality from the point of view of differential geometry, we start from an ideal situation that the mortality data set correspond to a smooth surface $\Sigma_s$ which is called a smooth mortality surface and is defined by a smooth function $z=f(t,x)$, where $t$ denotes time, $x$ denotes age and $z$ is the mortality of the people at the age of $x$ when time is $t$.

The mortality data for people born in year $t$ represent a curve $l_t$ on $\Sigma_s$ and we call the curve a cohort curve. If the cohort effect appears for these people, the mortality of this group has general characters and is distinguished from the mortality data of groups nearby. For this reason, at any $p\in l_t$, the oscillations along any other directions should be more observable than along the tangent vector of $l_t$. In differential geometry, an effective method to measure this kind of oscillation is to compute the normal curvature of a curve on a surface. The cohort curve with cohort effect should be a curvature line and the tangent vector of the curve should be one of the principle directions (Figure 1).

\begin{figure}[tbh]
\begin{centering}
\includegraphics[height=150pt,width=330pt]{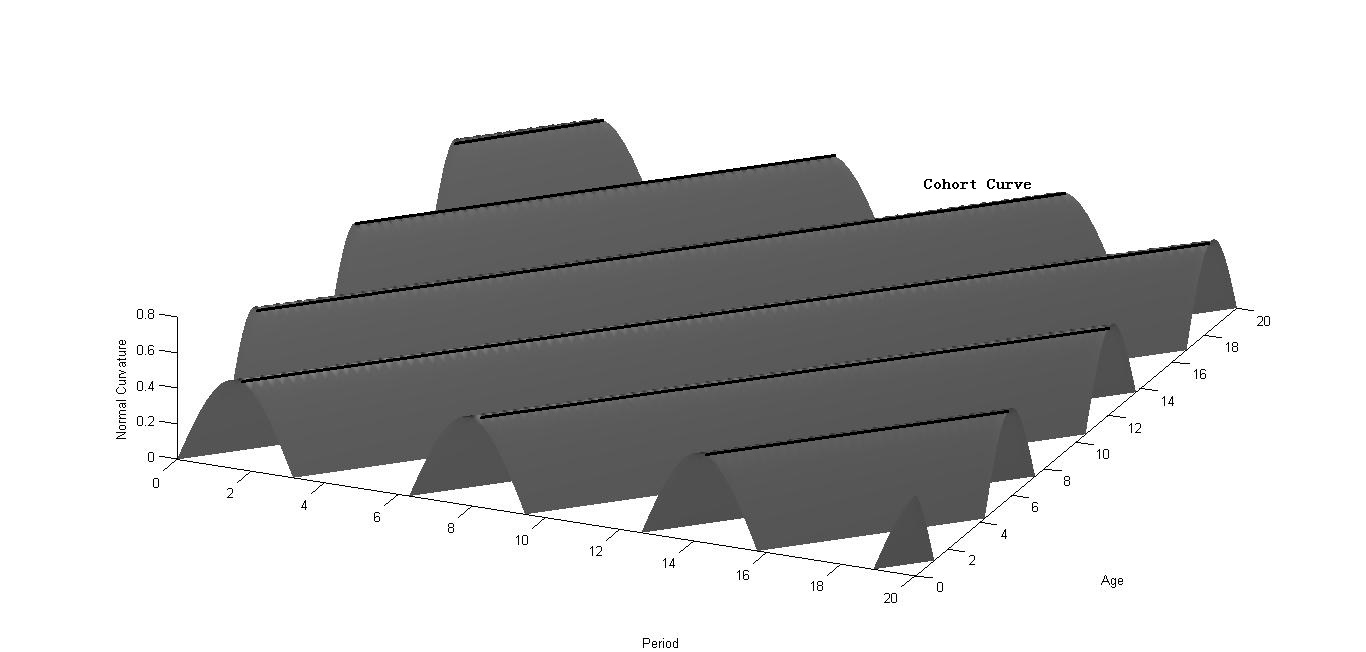}
\end{centering}
\caption[uk]{The sample of cohort curve which is not real,  just for explanation}
\end{figure}

But, in fact, since there are many kinds of noises of the mortality data and the data are discrete, we can not expect the cohort curve with cohort effect to be identical to a curvature line. For this reason, we define the cohort effect index mathematically as follow.\\

\noindent{\bf Definition:} {\it For a cohort curve on $\Sigma_s$, suppose $T(s)$ is the tangent vector of a point $p(s)\in l_t$ and $N(s)$ is the orthogonal direction of $T(s)$ on the tangent plane of $p(s)$. Let $NC_T (s)$ and $NC_N (s)$ be the normal curvatures along the directions $T(s)$ and $N(s)$ respectively. Then the integral
\be\label{index1}
CEI_t=\int_a^b |NC_T(s)-NC_N(s)|ds
\ee
is called the cohort effect index(CEI) of the generation born in year $t$. Here, for simplicity, we use the arc length $s$ as parameter to describe the cohort curve $l_t$ and the integrating range $[a,b]$ is decided by the structure of the mortality data.}\\

The reality is that the data of mortality are always discrete points on the surface $\Sigma_s$. We view the set of the points as a piecewise mortality surface and denote it by $\Sigma_p$ (Figure 2). Now we briefly describe how to realize the idea above on this piecewise surface $\Sigma_p$.

To begin our program, first we define the discrete parameter of a discrete curve $l$ contains three point $p_0$, $p_1$ and $p_2$. The discrete parameter is defined by
$$s_0=0,\ \ s_1=\frac{|p_1-p_0|}{|p_{1}-p_{0}|+|p_2-p_1|},\ \
s_2=1.$$
Next we estimate the tangent vector of $l$ at $p_1$. We call it a discrete tangent vector and denote it by $\vec{T}=(T_t, T_x, T_z)$. By minimizing the sum of the distances between the tangent line and the two points $p_0$ and $p_1$ under the constrain that the tangent line should pass through the point $p_1$, we can get an approximation of $\vec{T}$.
\be\label{tv1}
T_t=\frac{(s_0-s_1)\left(t(s_0)-t(s_1)\right)
+(s_2-s_1)\left(t(s_2)-t(s_1)\right)}{(s_0-s_1)^2+(s_2-s_1)^2},
\ee
\be\label{tv2}
T_x=\frac{(s_0-s_1)\left(x(s_0)-x(s_1)\right)
+(s_2-s_1)\left(x(s_2)-x(s_1)\right)}{(s_0-s_1)^2+(s_2-s_1)^2},
\ee
\be\label{tv3}
T_z=\frac{(s_0-s_1)\left(z(s_0)-z(s_1)\right)
+(s_2-s_1)\left(z(s_2)-z(s_1)\right)}{(s_0-s_1)^2+(s_2-s_1)^2}.
\ee

For a piecewise mortality surface $\Sigma_p$, We use $p_{ij}$ to denote points on the surface, where the subscript $i$ refer to year and the subscript $j$ refer to age and the coordinates of the point $p_{ij}$ in $\mathbb{R}^3$ are $(t_i,x_j,f(t_i,x_j))$. Since curvature of a curve $l_t$ at certain point $p_{ij}\in l_t$ is a local quality, to define the normal curvature on the discrete surface, it is appropriate to use only the points around $p_{ij}$ to compute the curvature. The minimal neighbourhood of $p_{ij}$ is consisting of the nine points $\{p_{i-1,j+1},p_{i,j+1},p_{i+1,j+1},p_{i-1,j},p_{ij},p_{i+1,j},
p_{i-1,j-1}, p_{i,j-1},p_{i+1,j+1}\}$. Among them, we view $l_1:\{p_{i-1,j-1},p_{ij},p_{i+1,j+1}\}$ and $l_2:\{p_{i-1,j+1},p_{ij},p_{i+1,j-1}\}$ as two discrete short curves whose tangent directions are taken as approximations of $T$ and $N$ in our definition of CEI.

By theories of differential geometry, for a smooth curve $l$ parameterized by $s$, suppose the unit tangent vector field along $l(s)$ is $\vec{V}(s)$,  then the curvature vector of the curve is defined by
\be\label{cv}
\vec{CV}(s)=\frac{\vec{V}'(s)}{|l'(s)|},
\ee
where $'$ is the derivative with respect to the parameter $s$. For the two discrete curve $l_1$ and $l_2$, by formulas (\ref{tv1}-\ref{tv3}) and normalization, we can get the unit discrete tangent vectors to $l_1$ and $l_2$ at point $p_{ij}$ and we denote them by $\vec{V}_1(p_{ij})=(v_{1t}(p_{ij}),v_{1x}(p_{ij}), v_{1z}(p_{ij}))$ and
$\vec{V}_2(p_{ij})=(v_{2t}(p_{ij}),v_{2x}(p_{ij}), v_{2z}(p_{ij}))$.
For a discrete curve, the derivative with respect to its discrete parameter can be defined by solving a similar constrained minimization problem as we do in estimating $\vec{T}$. Thus we can get the two discrete curvature vector fields $\vec{CV}_1$ and $\vec{CV}_2$ just following the formula (\ref{cv}).

Obviously, two unit tangent vectors $\vec{V}_1$ and $\vec{V}_2$ are not enough to determine a unique vector orthogonal to them. To get the normal vector of the surface $\Sigma_d$ at any point $p_{ij}$, we consider two more discrete curves across $p_{ij}$. Let $l_3:=\{p_{i-1,j},p_{ij},p_{i+1,j}\}$ and $l_4:=\{p_{i,j+1},p_{ij},p_{i+1,j-1}\}$, the same as we do for $l_1$ and $l_2$, we can get two unit tangent vectors $\vec{V}_3$ and $\vec{V}_4$. Since normal vector are orthogonal to any tangent vector, we can estimate the discrete unit normal vector $\vec{N}(p_{ij})$ by minimizing
$$f(\vec{N})=\sum_{k=1}^{4} |\vec{N}\cdot \vec{V}_k|^2,$$
with the constraint $\vec{N}\cdot \vec{N}=1$. For details to solve the problem, one can refer to \cite{19}. Finally, it is nature to define the discrete normal curvature along direction $\vec{V}_k$ at point $p_{ij}$ by
$$NC_k(p_{ij})=N(p_{ij})\cdot \vec{CV}_k(p_{ij}), \ \ k=1,2,3,4$$

For a fixed integer $m$, all the points $p_{ij}$ satisfying $i+j=m$ make up a curve related to persons born in the same year. We call these persons cohort $m$,or $C_{m}$ and call the curve a cohort curve. The tangent vector field along the cohort curve corresponds to $\vec{V}_1$ and we call this direction a cohort direction. By our definition of CEI for the smooth case, we define the discrete CEI for $C_{m}$ by
$$CEI_{m}=\sum_{i+j=m}|NC_1(p_{ij})-NC_2(p_{ij})|.$$
Now for any integer $m$ satisfying $a\leq m\leq b$, we get $CEI_m$. All these cohort effect indexes form a time series, and we call it the series of cohort effect (index) in the following of this paper.

\begin{figure}
\includegraphics[height=120pt,width=300pt]{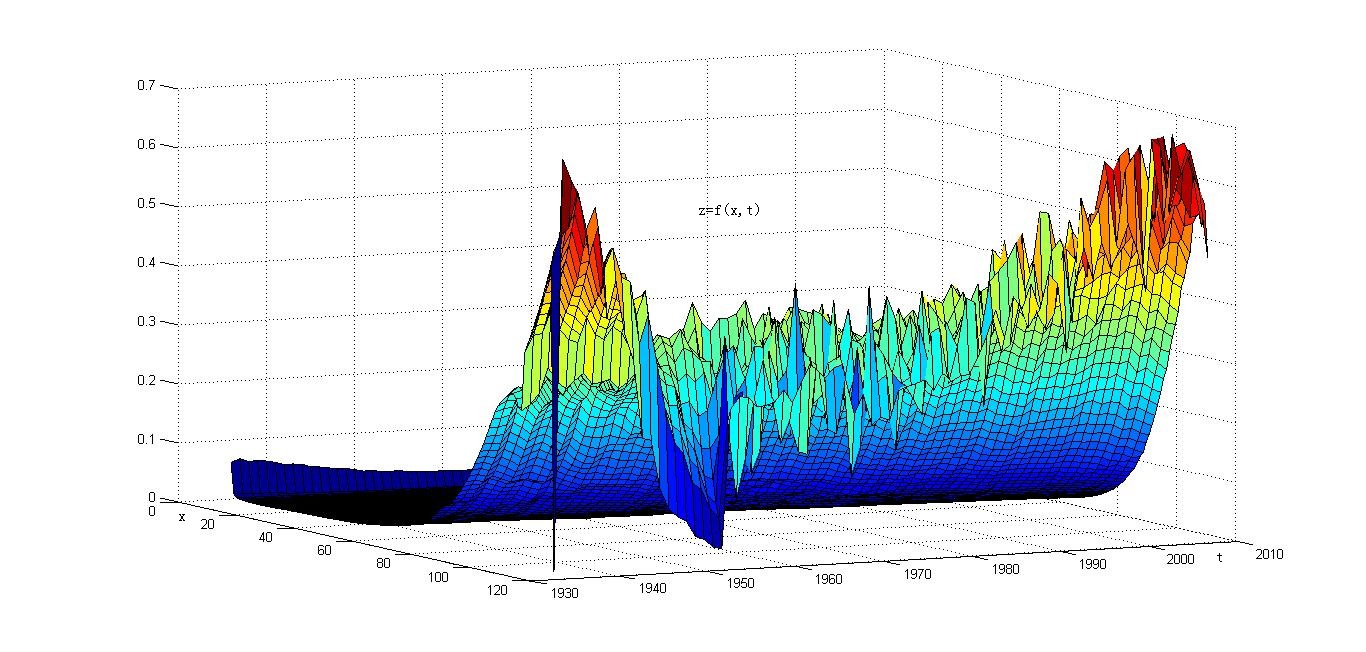}
\caption[surface]{An example of the mortality surface}
\end{figure}

\section{Applications of the model to identify and measure cohort effect}

    Our data come from the "The Human Mortality Database" \footnote{http://www.mortality.org/}. Several types of data sets can be used. We choose the data sets of "Death rate" and "1 $\times$ 1" for our practice and we also use the data sets of "1 $\times $ 5" and "5 $\times $ 5" for auxiliary check or comparison.

\begin{figure}[tbh]
\begin{centering}
\subfloat[first]{\begin{centering}
\includegraphics[height=90pt,width=185pt]{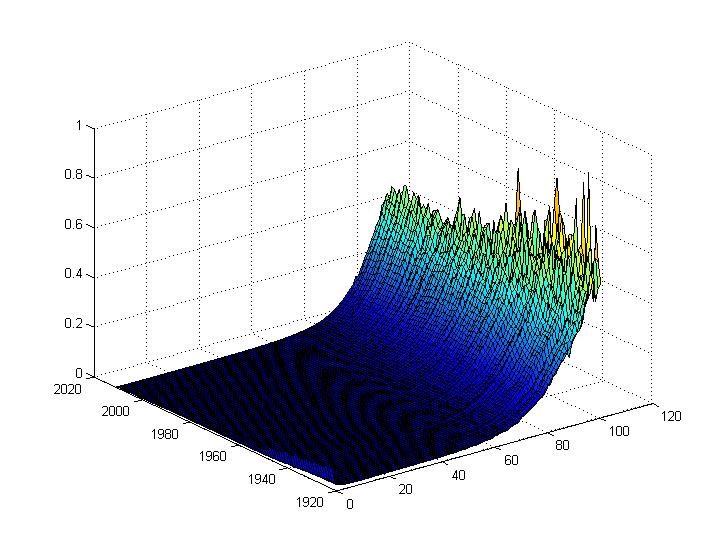}
\par \end{centering}
}
\subfloat[second]{\begin{centering}
\includegraphics[width=185pt]{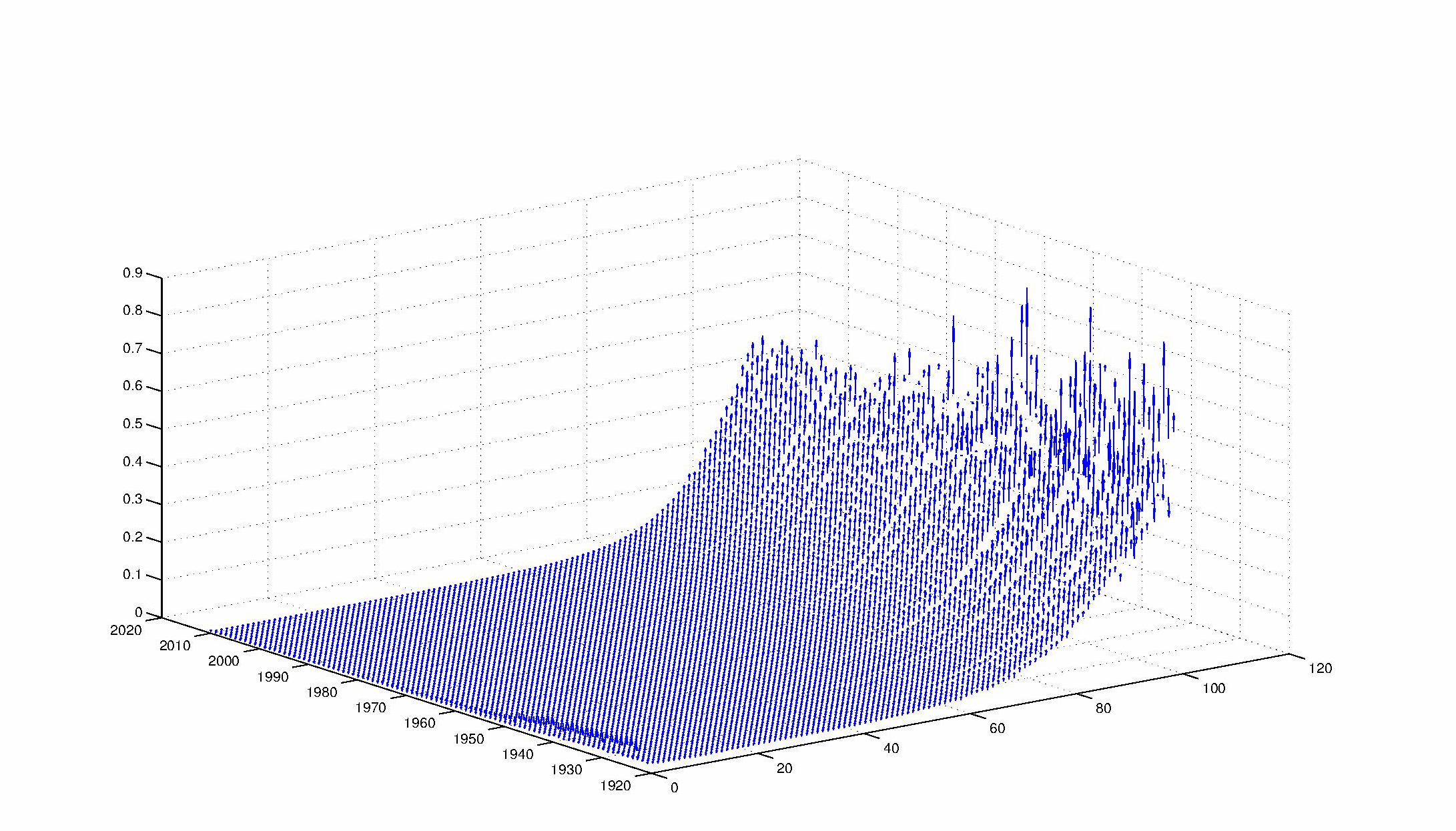}
\par\end{centering}
}

\subfloat[third]{\begin{centering}
\includegraphics[width=185pt]{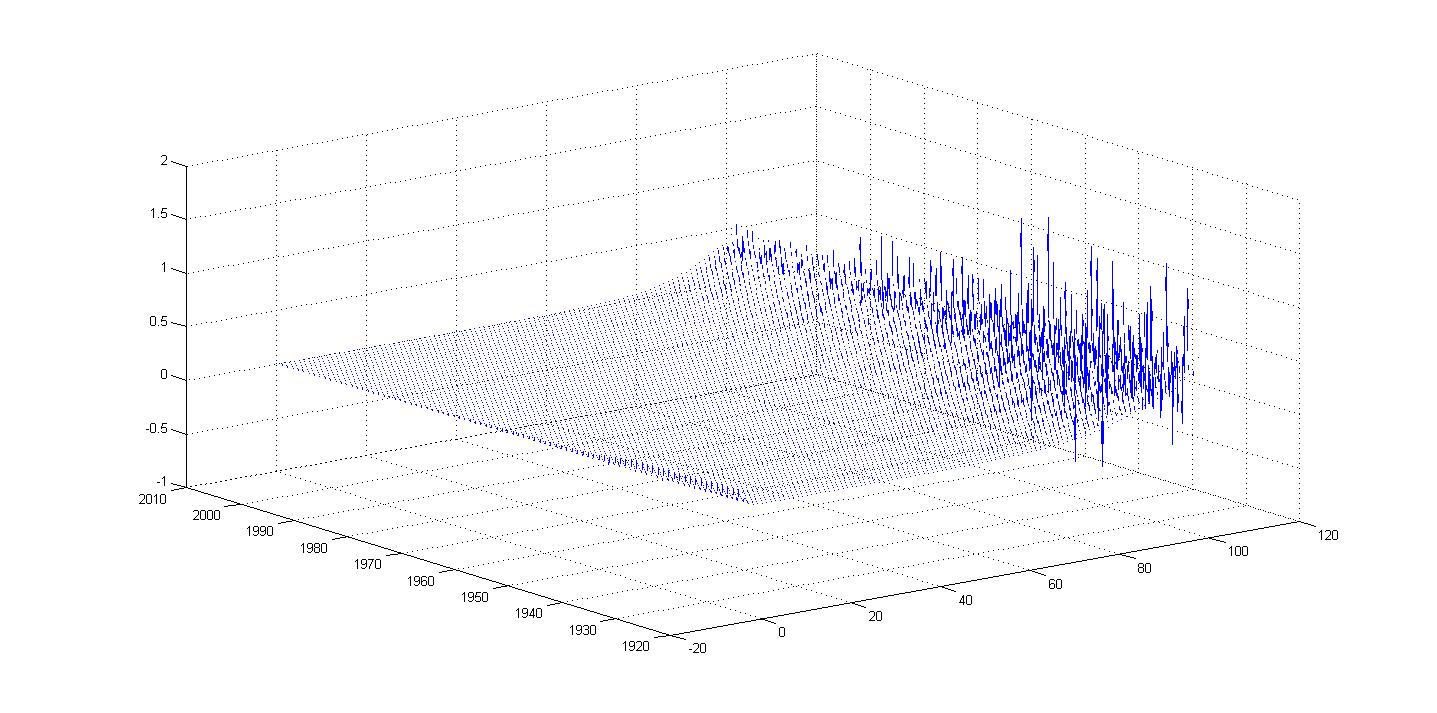}
\par\end{centering}
}
\subfloat[fourth]{\begin{centering}
\includegraphics[width=185pt]{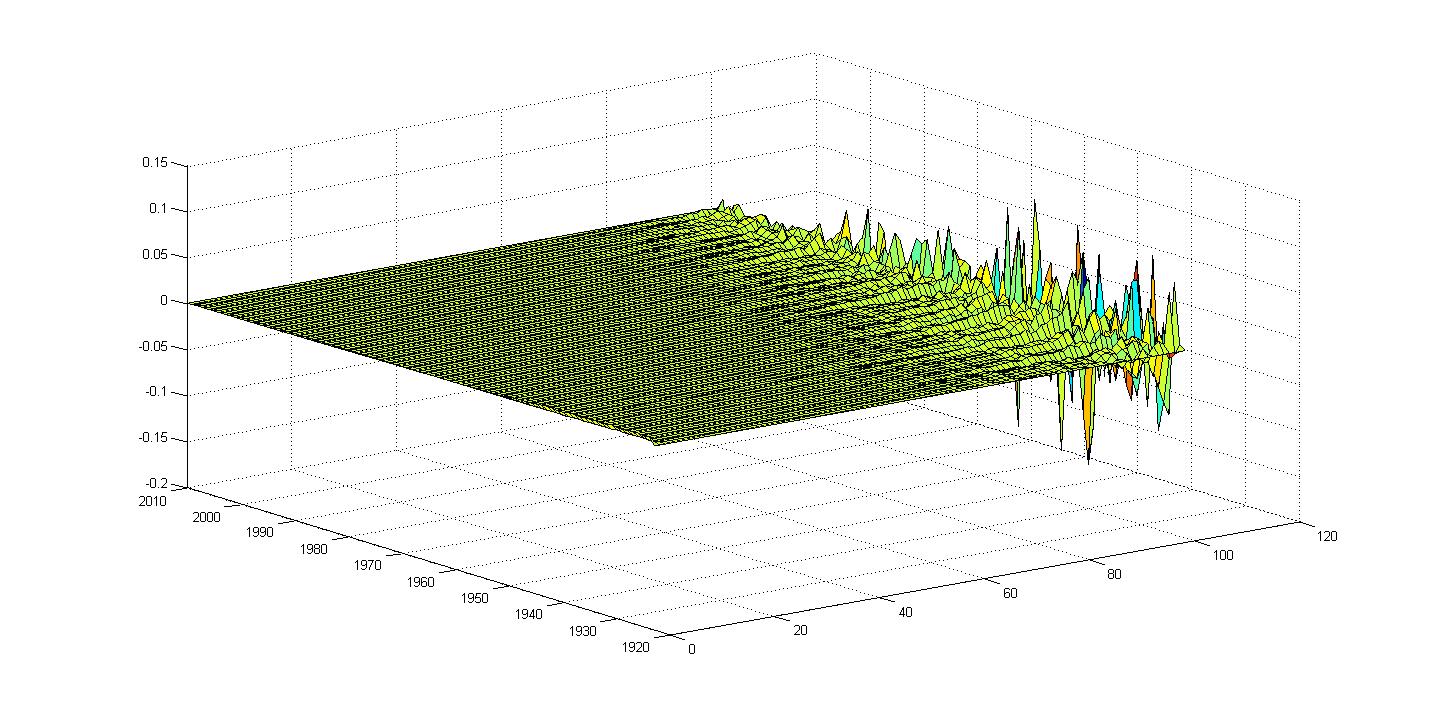}
\par\end{centering}
}

\par\end{centering}
\caption[first]{The process of calculating the series of cohort effect}
\end{figure}

\begin{figure}[tbh]
\begin{centering}
\subfloat[The cohort effect series of U.K.]{\begin{centering}
\includegraphics[height=90pt,width=185pt]{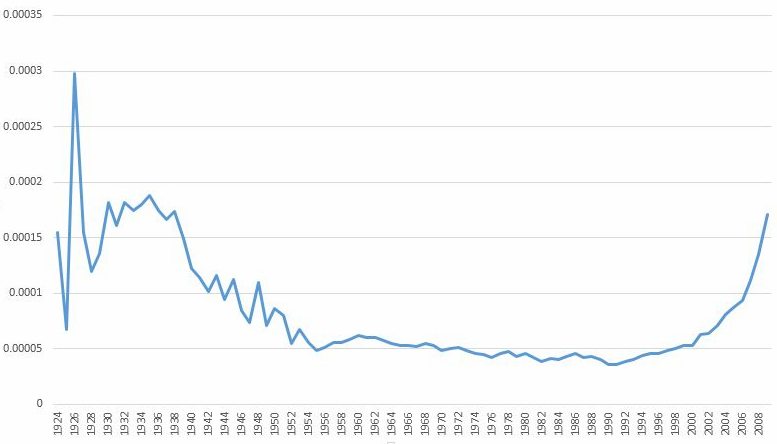}
\par \end{centering}
}
\subfloat[3-dimensional block graph]{\begin{centering}
\includegraphics[height=90pt,width=170pt]{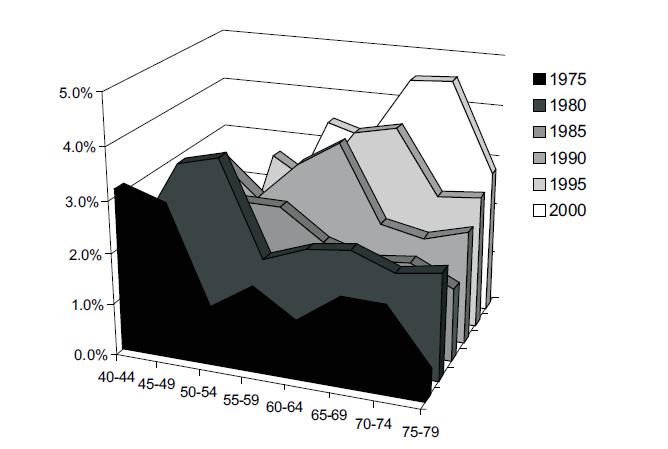}
\par \end{centering}
}
\end{centering}
\caption[uk]{The cohort effect of U.K.}
\end{figure}

  We first practise this model in the mortality data of U.K.. Its cohort effect of birth year is frequently referred to in references. The four figures in Figure 3 showed the calculating process intuitively. The original mortality data set corresponds to a discrete mortality surface, showed by the upper left figure (a). The results of calculating tangent vectors and curvature vectors are four vector fields. The upper right figure (b) and the left lower figure (c) in the sketch diagram the results along the first direction (cohort effect direction we concerned). Finally, the normal curvature along the direction of the cohort effect can be showed as another surface which is the right lower figure (d).

  Although we can get some direct impression about the mortality data from the surface of normal curvature, it is still difficult to recognize cohort effect exactly. Just as what has been described in our model, the detection can be carried out on the cohort effect series in the 2-D plane. Every peak of the series means that there exists obvious cohort effect since the mortality of this generation is apparently different from those of the neighbors.

   The left figure in Figure 4 gives the cohort effect series of U.K. which just comes up to our expectations. We can find two obvious peaks there. One peak is about $3\sim 5$ years, and another peak is about $10$ years which include several small peaks. The two peaks show strong birth-year cohort effects. This cohort effect series is in accord with the well-known U.K. cohort effect which has been noted several times in literatures. For example, the right figure in Figure 4 which comes from Willets \footnote{R.C. Willets, The cohort effect: insights and explanations, 2004, Page 2} uses a three-dimensional block graph to show cohort effect of U.K. intuitively. The two graphs above both give the conclusion that the generations (of both sexes)who born approximately between 1925 and 1945 have experienced more rapid mortality changing.

   Moreover, more detailed information about cohort effect could be seen intuitively in our result ((a)in Figure 4):

   1)  By the theory of our model, the generation born in 1925 holds obvious mortality cohort effect since there is a highest peak in the final series. The reasons for that are not precisely understood. Maybe just data errors or outliers dominate it. For cohort effect detection, just one year is too short to represent one generation in view of scale.

   2) The generation born between 1930 and 1935 holds another obvious mortality cohort effect. The corresponding series is approximately a plateau but not a peak. $6$-year is enough to represent a generation and we view this generation as the dominating part of cohort effect mentioned in literatures. The reason that we can reduce the gap in literatures from $[1925,1945]$ to $[1930-1935]$ is nothing but calculation.

\begin{figure}[tbh]
\begin{centering}
\subfloat[U.K.]{\begin{centering}
\includegraphics[height=80pt,width=185pt]{ukfig.jpg}
\par \end{centering}
}
\subfloat[U.S.]{\begin{centering}
\includegraphics[width=185pt]{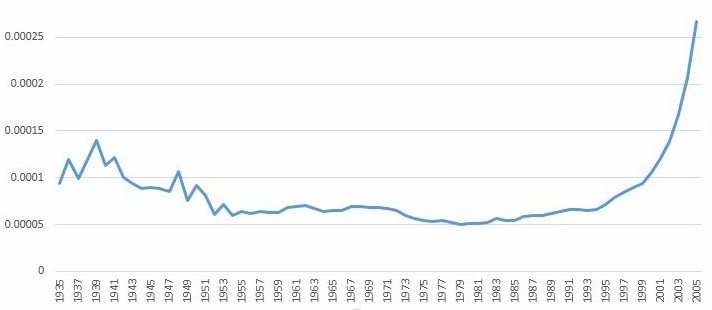}
\par \end{centering}
}

\subfloat[Canada]{\begin{centering}
\includegraphics[width=185pt]{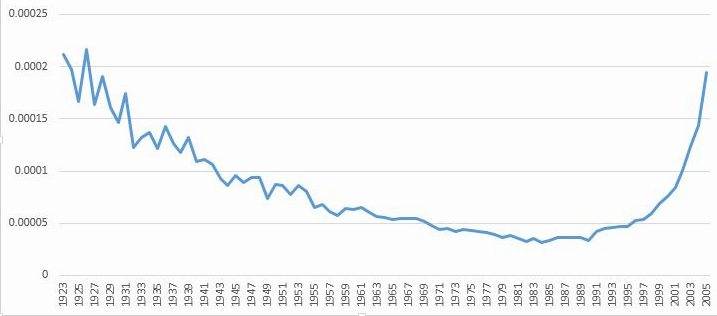}
\par \end{centering}
}
\subfloat[Japan]{\begin{centering}
\includegraphics[width=185pt]{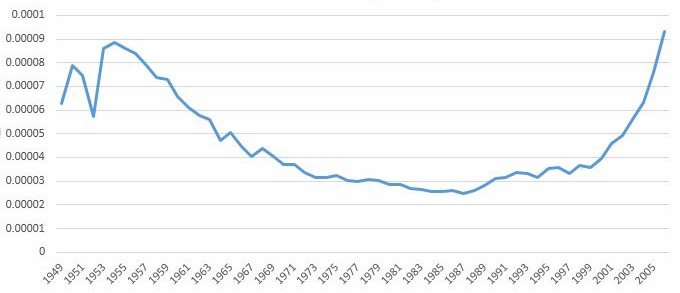}
\par \end{centering}
}
\end{centering}
\caption[uk]{series of cohort effect }
\end{figure}

     We also apply our method to the data sets of other three countries and the series of cohort effect are showed in Figure 5. The upper left one is the series of U.K., the upper right one is the series of U.S, the left down one is the series of Canada and the right down one is the series of Japan. By the method we used above, we assert that, for the series of U.S., cohort effects exist in the gap [1933,1941] and the gap [1948,1953] and there is no other obvious cohort effect. The situation similar to U.S. exists on the series of Japan, but the gaps are [1949,1952] and [1953,1958]. The volatility of the series of Canada is more drastic and this means that mortality of neighboring generations are much more different.

    There is one common ground that all of the four series have a stable upward trend and then a sharp drop at the end. Here we provide an explanation for this kind of behavior of the series.

    1) The upward trend is the consequence of shorter data. There is fewer and fewer data along the same cohort direction, for example, only one data in the last birth year (age 0 and calender year 2010). This leads to much volatility in calculating normal vectors based on discrete algorithm of differential geometry. So the oscillations of mortality for younger generations are bigger than those for the older generations.  Since this definitely brings errors for measuring cohort effects, we should just consider the former part of the series of cohort effect and remove the tail of the youngest generations. In fact, it is unpersuasive to claim any cohort effect when we capture little information about their mortality.  We recommend to detect and measure the cohort effect for the generations who born before at least 1970.

    2) The cause of sharp drop in the tail is because of our initial conditions for differential calculation. We set the all border data to zero to keep the same size with original data sets in the process of differential calculation.

    {\bf Remark:} As what we have just explained, the sharp drops of the series are useless when we deal with cohort effects, so we cut off the tails of the series in our figures of series of cohort effect.

    We find that "U" shape is a general result when we apply our method on the data sets of more countries. "U" shape means that the generations who born between 1950 and 1980 experience stable mortality improvement. Furthermore, the mortality volatility tends to decrease since 1940 in most of countries. It is convinced that the post-war prosperity of most countries contributes to "U" shape.

\begin{figure}[tbh]
\begin{centering}
\includegraphics[height=120pt,width=300pt]{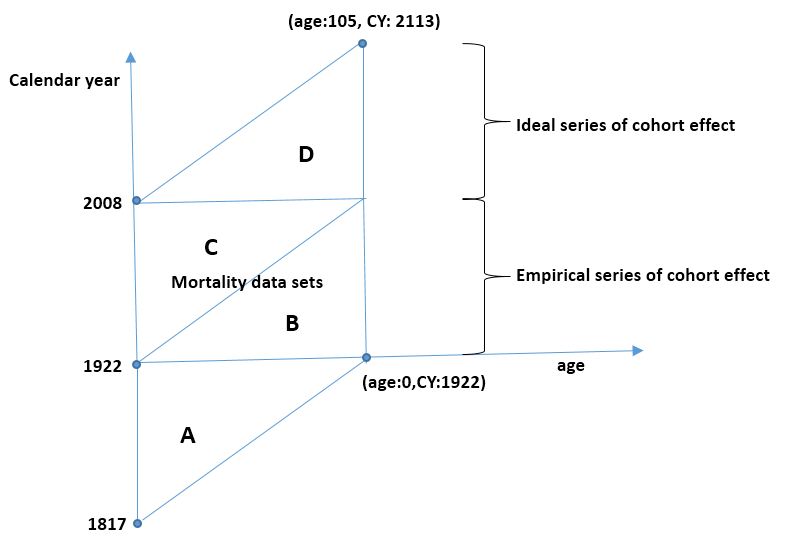}
\end{centering}
\caption[dataset]{The data sets and series of cohort effect}
\end{figure}

\begin{figure}[tbh]
\begin{centering}
\includegraphics[height=100pt,width=300pt]{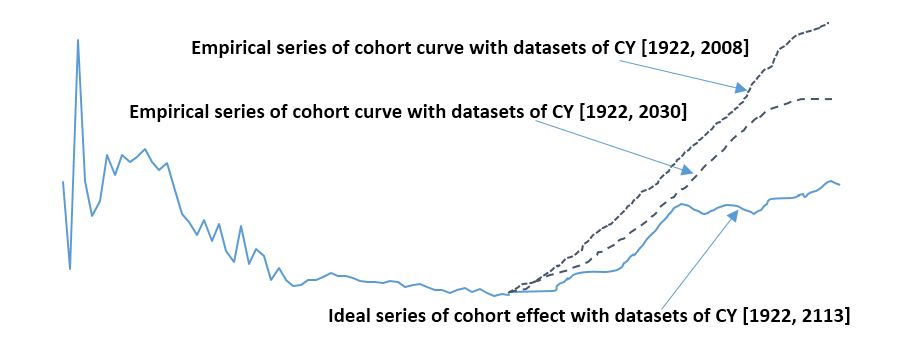}
\end{centering}
\caption[dataset]{Ideal and empirical series of cohort effect}
\end{figure}

     Just like what we have explained above, "U" shape is related to the length of data sets what we use. Here we make a formal description with the auxiliary help of Figure 6 and Figure 7. If the data sets are complete, we can get an ideal series of cohort effect by using all need data which includes Part A, Part B, Part C and Part D (Figure 6).  Taking U.K. as an example, if we track the generations from 1922 to 2008, the mortality at age 105 in calendar year 2113 is needed (When maximal age is 105) but this is impossible. And the usual mortality data set is a data matrix which consists of Part B and  Part C, as shown in Figure 6. Since we can not track all the targeted generations from their birth to death, we only work on Part C of the data set. By this way, we get an empirical series of cohort effect instead of an ideal series of cohort effect. So the above series of cohort effect in Figure 5 (U.K., US, Canada and Japan) are the empirical series of cohort effect.

     The series which is calculated from part C and D of the data set are called the ideal series of cohort effect. There is a direct relationship between ideal and empirical ones. When a new row of data (for example, the mortality data for all ages in 2009) is added into the data set of Part C, the new empirical series of cohort effect will be closer to the ideal series of cohort effect. The ideal series of cohort effect shows the accurate information about cohort effect but it is impossible to obtain. Since the empirical series of cohort effect is a good approximation of the ideal series of cohort effect especially for the older generations, we can use the empirical series of cohort effect to find desired information.

     So for the "U" shape in the empirical series of cohort effect, the left part of 'U' can provide us desired information about cohort effect and the right part of 'U' does not provide any information but comes from the insufficiency of the data set.

\section{Aggregating index of cohort effect }

  In actuarial research, it is significative to quantify the strength of cohort effect for the whole population. In particular, when we make a choice of mortality models, we can decide whether or not to take cohort effect into account according to the strength of the effect in this data set. So we would like to construct an index to measure the strength for the whole population which may be called the aggregating index of cohort effect (AICE). Although some potential mathematical theories seem to contribute to this purpose, we had better find a simple statistics firstly.

  {\bf Definition: } {\it For any empirical series of cohort effect, the aggregating index of cohort effect is defined as its coefficient of variation, in other words, the quotient of its sample standard deviation divided by its sample mean.}

  The strength of the cohort effect for the whole population should consider two factors: the average difference between neighboring generations and the average overplus of the outstanding generations. The standard deviation gives a measurement how far the series of cohort effect spreads out and the mean gives a whole description for the series of cohort effect.

  Here we make an explanation for its rationality. For the time series of the cohort effect, some most-frequently used statistics are listed in Table 1. During the computation of these results, we only use the series between $[1922,1970]$ instead of the whole series of cohort effect to reduce the effect of missing data in Part C (Figure 6). From this table, the coefficient of variation does well in serving for measuring cohort effect for the whole population. The larger the coefficient of variation is, the more obvious the cohort effect is. The maximal value of the coefficient of variation in Table 1 is $0.6177$ for UK and this means that UK has the most obvious cohort effect in these four countries. This is consistent with many literatures intuitively. Although the variance or standard deviation seem to do the same, but there are two problems we have found:

  1) The final value should be adjusted to be more like an index. In the Table 1, "E-5" or "E-09" is not convenient for application.

  2) Without considering the average difference between neighboring generations, the results may be in contradiction with the results by direct observation of mortality data.

   Another derivative parameter from series of cohort effect is the generation gap (or cohort-effect generation gap) which describes how long cohort effect maintains. Formally, it is the gap from the beginning to the end of a peak on the series of cohort effect.  So there are many generation gaps in view of cohort effect on the ideal series of cohort effect. And we recommend to consider only the part before 1970 when using empirical series of cohort effect.

   Usually, in social sciences, we use $5$ or $10$ years to represent a generation. But our results show that the length of a generation is a problem in itself and we give a method to resolve the problem. Table 2 gives the maximal and minimal generation gaps for the four targeted countries. This is also an important character for the population.

\begin{table}
\caption{Aggregation index of cohort effect}
\begin{tabular}{|r|r|r|r|r|}
\hline
           &         US &     Canada &      Japan &         UK \\
\hline
      mean & 8.14646E-05 & 8.31718E-05 & 4.65899E-05 & 8.24972E-05 \\
\hline
  Variance & 1.32031E-09 & 2.31223E-09 & 4.04465E-10 & 2.59693E-09 \\
\hline
     stdev & 3.63361E-05 & 4.80857E-05 & 2.01113E-05 & 5.09601E-05 \\
\hline
   coefficient of variation & 0.446035861 & 0.578148277 & 0.431667221 & 0.617719261 \\
\hline
\end{tabular}
\end{table}

\begin{table}
\caption{The cohort-effect generation gap}
\begin{tabular}{|r|r|r|}
\hline
           & Minimal generation gap (years) & Maximal generation gap (years) \\
\hline
United Kingdoms &          3 &         10 \\
\hline
United States &          2 &          5 \\
\hline
    Canada &        1.5 &          3 \\
\hline
     Japan &          2 &          4 \\
\hline
\end{tabular}
\end{table}

\section{Conclusion and the future work}

   We promote an effective method based on differential geometry to implement quantitative measurement of cohort effect. The peaks on the series of cohort effect mean the existence of cohort effects and the height of the peaks tells us the strength of the corresponding effects. We also apply this method on the data sets of four countries including the United Kingdom, the United states, Canada and Japan. All the resulting series show the desired strength of cohort effects in different generations. In particular for U.K., our method can give a further description of the well-known mortality cohort effect. Based on the series of cohort effect, we introduced the aggregating index of cohort effect (AICE) which is a general description of cohort effect for the whole population of a country or group.

   Applications of our model in the analysis on longevity risk are one of the problems we are considering. Since the series of cohort effect measure the strength of cohort effect for different generations, we can introduce a parameter into classical mortality models: for example, Lee-Carter model may be changed into $\ln\mu_x(t)=\alpha _x +\beta_x k_t+\sigma c_{i-x}$, where $c_{i-x}$ is decided by the series of cohort effect. The details are ongoing and we have got some elementary results. Statistical work of AICE for different countries and its applications in longevity risk are another problem which is worthy of attention. Besides, further studies on the theoretical model are in progress. We hope to make the model subtler and more intelligent and analyze the contributing factors of the detected cohort effects.

   Finally, we must repeat a fact in the series of cohort effect of U.K. that many ups and downs exist between the gap of [1925,1945]. It means that the difference among the generations is detected and this may help us to understand or to explain some characters of the population. We believe that a number of explanatory factors remain to be found.

{\bf Acknowledgement.}
 The first author is partially supported by the MOE Project of Key Research Institute of Humanities and Social Sciences at Universities (11JJD790004), JIAOBAO Fund from the Insurance Institute of CHINA (jiaobao2013-03) . The second author is partially supported by NSFC 11201028 and the Fundamental Research Funds for the Central Universities. Both authors thank the support of Data Lighthouse Plan.

\end{document}